\documentclass[preprint,1p,authoryear]{elsarticle}
\usepackage{amsmath}
\usepackage{graphicx}

\begin{document}

\begin{frontmatter}

\title{Constraints on dark energy from the lookback time versus redshift test }

\author[ksu,gno]{Lado Samushia}\ead{lado@phys.ksu.edu}
\author[ud]{Abha Dev}\ead{abha\_dev@yahoo.com}
\author[ddu]{Deepak Jain}\ead{djain@ddu.du.ac.in}
\author[ksu]{Bharat Ratra}\ead{ratra@phys.ksu.edu}

\address[ksu]{Department of Physics, Kansas State University, 116 Cardwell Hall, Manhattan, KS 66506, USA.}
\address[gno]{National Abastumani Astrophysical Observatory, Ilia Chavchavadze University, 2A Kazbegi Ave, Tbilisi GE-0160, Georgia.} 
\address[ud]{Miranda House, University of Delhi, Delhi 110 007, India.}
\address[ddu]{Deen Dayal Upadhyaya College, University of Delhi, New Delhi 110 015, India.}

\begin{abstract}
We use lookback time versus redshift data from galaxy clusters \citep{cap}
and passively evolving galaxies \citep{svj}, and apply a bayesian prior on the total age of the Universe based on
WMAP measurements, to constrain dark energy
cosmological model parameters. Current lookback time data provide interesting and moderately
restrictive constraints on cosmological parameters. When used jointly with current baryon
acoustic peak and Type Ia supernovae apparent magnitude versus redshift data, lookback time data tighten the constraints
 on parameters and favor slightly smaller values of the nonrelativistic matter energy density.
\end{abstract}

\begin{keyword}
cosmological parameters \sep galaxies: general \sep galaxies: clusters: general
\PACS 95.35.+d \sep 98.80.-k
\end{keyword}

\end{frontmatter}

\section{Introduction}
It is now a well established fact that the expansion of the Universe is
accelerating, but the underlying mechanism which gives rise to this
cosmic acceleration is still a mystery. Recent cosmological
observations including the Hubble diagram of Type Ia supernovae \citep[SNeIa, e.g.,][]{hicken09,schafieloo09,guimaraes09},
combined with cosmic microwave background (CMB) anisotropy
measurements \citep[e.g.,][]{dunkley09,komatsu08}, baryon acoustic peak galaxy power spectrum data \citep[e.g.,][]{percivaletal07,samushia09a,gaztanaga08,wang09}, 
and galaxy cluster gas mass fraction measurements
\citep[e.g.,][]{allen08,samushia08,ettori09} 
indicate that we live in a spatially-flat universe where nonrelativistic matter contributes about $30\%$ of
the critical density. Within the framework of
Einstein's general theory of relativity, the rest of the $70\%$ of the 
energy density of the Universe is termed dark energy, a
mysterious component with negative effective pressure that is responsible for the
observed accelerated expansion.\footnote{For discussions of modification
of Einsteinian gravity on cosmological scales that attempt to do away with the 
need for dark energy, see \citet{rapetti09}, \citet{bamba09}, \citet{capozziello09}, \citet{wu09}, \citet{zhan09},
and references therein.} For recent reviews of dark energy see \citet{ratra08}, \citet{caldwell09}, \citet{frieman09}, and \citet{sami09}.

There are many dark energy candidates. The 
simplest is Einstein's cosmological constant $\Lambda$. In addition, there are
other options like XCDM, a slowly rolling scalar field,
Chaplygin gas, etc., which can also give rise to an accelerated expansion of
the Universe. In this paper we constrain the parameters of three
different dark  energy models. The first model is the 
cosmological constant  dominated cold dark matter ($\Lambda$CDM)
model \citep{pee84}.In this model the energy density of the vacuum (the  cosmological
constant) does not vary with time and it has a negative
pressure characterized by $p_{\Lambda} \, = \,- \rho_{\Lambda}$, where
$\rho_{\Lambda}$ is the vacuum energy density.

Secondly, we consider the XCDM parameterization of dark
energy. In this case dark energy is assumed to be a fluid  satisfying
the following relation between pressure and the energy density, $p_{\rm x} =
\omega_{\rm x}\rho_{\rm x}$, with $\omega_{\rm x}<0$; this is not a physically complete model. 
Lastly, we study the slowly rolling dark energy scalar
field $\phi$ model ($\phi$CDM) with an inverse power-law potential energy density for
the scalar field, $V \propto \phi^{-\alpha}$ where $\alpha$ is a nonnegative constant
\citep{pr,rp, pr03}.\footnote{In the $\phi$CDM model we consider here, $\phi$ only couples gravitationally to other components. For models where $\phi$ also interacts more directly with other components, see \citet{chen09}, \citet{baldi09}, \citet{bento09}, \citet{pettorino09}, \citet{guvela09}, \citet{lavacca09}, and references therein. For other dark energy models,
see \citet{lu09}, \citet{arbey09}, \citet{hrycyna09}, \citet{basilakos09},
\citet{tsujikawa09}, \citet{dutta09}, \citet{neupane09}, and references
therein.} We only consider the spatially-flat $\phi$CDM and XCDM cases. The $\phi$CDM model with $\alpha = 0$ and the XCDM model
with $\omega_{\rm x} = -1$ are equivalent to the spatially-flat $\Lambda$CDM
model with the same matter density. In all three models the
nonrelativistic matter density is dominated by cold dark matter.   

In this paper we use two sets of lookback time versus redshift measurements,
for galaxy clusters \citep{cap} and for passively evolving galaxies \citep{svj},
and apply a bayesian prior on the total age of the Universe based on WMAP estimates \citep{dunkley09},
to constrain parameters of these dark energy
models. This time-based cosmological test differs from other widely-used distance-based cosmological tests.\footnote{Distance-based cosmological tests include those 
mentioned above that use SNeIa, CMB, baryon acoustic peak, and galaxy cluster gas mass fraction data,
as well as radio-galaxy and quasar angular size versus redshift data \citep[e.g.,][]{chen03,podariu04,daly09,santos08}
and gamma-ray burst luminosity distance versus redshift measurements \citep[e.g.,][]{tsutsui09,qi09,wei08,liang09,wang08,samushia09c}.}
An important
feature of this time-based method is that the age of distant objects
are independent  of each other. Therefore, it may avoid biases that are present
in techniques that use distances of primary or secondary
indicators in the cosmic distance ladder method. In the literature a variety of time-based
methods have been considered, based on measurements of the absolute age of 
objects, differential age of objects, and lookback time of objects.\footnote{A variation of this test
uses measurements of the Hubble parameter as a function of redshift \citep[e.g.,][and references therein]{samushia06,lin08,dev08,fernandez08}.}

The absolute age method is based on the simple criterion
that the 
age of the Universe at a given redshift is always greater than or
equal to the age of the oldest object at that redshift
\citep{a11,a12,dj1,wei07}. The differential age method is based on the measurement of $\Delta z/\Delta t$.
$\Delta z$ is the redshift separation between the two passively evolving galaxies
having the age difference $\Delta t$ \citep{jl,jetal}. This method
requires a large sample of passively evolving galaxies with high
quality spectroscopy and is probably more reliable than the absolute age method
as a number of systematic effects are eliminated. 

Lookback time as a tool to
constrain 
dark energy models was first used  by \citet{cap} who compiled a list of galaxy cluster ages and redshifts and used this
data to constrain the XCDM dark energy parameterization. This data has been used to constrain brane cosmology and 
holographic dark energy models \citep{pir,yi}. The lookback time test has also been applied using passively evolving galaxies data,
to constrain parameters of XCDM and $\Lambda$CDM \citep{dj,alc}.
No doubt these time-based methods are subject to some
different systematic errors but they offer an
independent means to cross-check cosmological constraints obtained using
other techniques. 

In this paper we take advantage of the fact that the \citet{cap}
galaxy cluster data and the \citet{svj} passive galaxy data are independent,
so it is straightforward to use them simultaneously in a lookback time versus
redshift test analysis of dark energy models. Our joint analyses of these data sets allow
us to derive the tightest lookback time constraints on dark energy parameters to date.
The resulting constraints are moderately restrictive, and these data favor lower
matter density values than do some other current data, but are consistent with
a spatially-flat $\Lambda$CDM model in which nonrelativistic matter
contributes 30\% of the energy budget at a little less than two standard
deviations.
To derive tighter constraints, we perform a joint analysis of the lookback time
data with current baryon acoustic peak and SNeIa measurements.

In Sec. 2 we describe the lookback time as a function of redshift
test. The data and method we use are outlined in Sec. 3.  Our results are presented and
discussed in Sec. 4.


\section{Lookback time versus redshift test }

The lookback time is the difference
between the present age of the Universe $ (t_0)$ and its
age at redshift $z$, $t(z)$,
\begin{eqnarray}
t_{L}(z, p) & = & t_0(p)\, - \, t(z)
 = {1\over H_0}\left[\;\int_0^{\infty}{dz'\over(1 +
z'){\cal{H}}(z',p)}\,\, - \,\, 
\;\int_z^{\infty}{dz'\over(1 + z'){\cal{H}}(z',p)}\right]\nonumber \\ 
&&\nonumber \\
& = &{1 \over H_0} \;\int_0^{z}{dz'\over(1 + z'){\cal{H}}(z',p)}.
\end{eqnarray}
\noindent Here $p$ are the 
parameters of the cosmological model under consideration, 
${\cal{H}}(z,p)= H(z,p)/H_0$, 
$H(z,p)$ is the Hubble parameter at redshift $z$, and the Hubble constant
$H_0=100h\ {\rm km\ s^{-1} Mpc^{-1}}$. 

Following \citet{cap}, the observed lookback time $t_{L}^{\rm obs}(z_i)$, to
an object $i$ at redshift $z_i$ is defined as
\begin{equation}
t_{L}^{\rm obs}(z_i,t_{\rm inc},t_0^{\rm obs})= t_0^{\rm obs}- t_i(z_i)-t_{\rm inc}.
\end{equation}
\noindent Here

\noindent $\bullet$ $t_0^{\rm obs}$ is the measured current age
of the Universe.  

\noindent $\bullet$ $t_i(z_i)$ is the age of the object (passively evolving galaxy, cluster,
etc.), defined as the difference between
the current age of the Universe at redshift $z_i$ and the age of the Universe when the object
was born at redshift $z_f$,

\begin{equation}
t_i(z_i)=t(z_i) - t(z_f) =  t_{L}(z_f) \, - \, t_{L}(z_i)
 =  {1\over H_0}\;\int_{z_{i}}^{z_f}{dz'\over(1 + z'){\cal {H}}(z',p)},
\end{equation}

\noindent where we have used Eq.~(1).

\noindent $\bullet$ $t_{\rm inc}=t_0^{\rm obs}-t_L({z_f})$ is the incubation time of the
object. This delay factor encodes our ignorance of the formation redshift
$z_f$.

To compute model predictions for the lookback time $t_L(z,p)$, Eq.~(1),
we need an expression for $H(z,p)$. In the $\Lambda$CDM model the Hubble parameter is
\begin{equation}
H(z,p)\, =H_0 \,[ \Omega_{\rm m}(1 +z)^3 + (1 - \Omega_{\rm m}
  -\Omega_{\Lambda})(1 +z)^2 + \Omega_{\Lambda}]^{1/2},
\end{equation}
\noindent where $p$ are $\Omega_{\rm m} $ and $\Omega_{\Lambda}$, the
nonrelativistic matter and dark energy density parameters at $z=0$. For the XCDM parameterization in a spatially-flat cosmological model we have 
\begin{equation} 
H(z,p)\, =H_0 \,[ \Omega_{\rm m}(1 +z)^3 + (1 - \Omega_{\rm m})(1
  +z)^{3(1+\omega_{\rm x})} ]^{1/2},  
\end{equation}
\noindent where $p$ are $\Omega_{\rm m}$ and $\omega_{\rm x}$.
\noindent In the spatially-flat $\phi$CDM model
\begin{equation}
H(z,p)\, =H_0 \,[ \Omega_{\rm m}(1 +z)^3 + \Omega_{\phi}(z)]^{1/2},  
\end{equation}
\noindent where the scalar field energy density parameter $\Omega_{\phi}(z)$
can be evaluated numerically by solving the coupled set of equations
of motion,
\begin{equation}
\ddot{\phi} + 3{\dot{a}\over a}{\dot{\phi}} - {\kappa\alpha \over 2}
m_p^{2} \phi^{-(\alpha+1)} = 0,
\end{equation}
\begin{equation}
\left({\dot{a}\over a }\right)^2\, = \, {8\pi \over {3\,m_p^2}}[ \Omega_{\rm m}(1 +z)^3 +
  \Omega_{\phi}(z)], 
\end{equation}

\begin{equation}
\Omega_{\phi}(z) \, = \, [ (\dot{\phi})^2 +
  \kappa m_p^2\phi^{-\alpha}]/12.
\end{equation}
\noindent Here $a(t)$ is the scale factor, an overdot denotes a time
derivative, $m_p$ is Planck's mass, and $\kappa$ and $\alpha$
are non-negative constants that characterize the inverse power law potential
energy density of the scalar field, $V(\phi)=\kappa\phi^{-\alpha}$. In this
case the parameters $p$ are $\Omega_{\rm m}$ and $\alpha$.
  
\section{Data and computation}  

In order to constrain cosmological parameters of $\Lambda$CDM, 
XCDM, and $\phi$CDM, we use two age data sets. One is the \citet{svj} ages of 32 passively evolving 
galaxies (Table~1, R.~Jimenez, private communication 2007) in the redshift interval $0.117
\le \,z \, \le 1.845$. For this sample we assume a $12\%$ one standard deviation uncertainty on the age
measurements (R.~Jimenez, private communication~2007). The other is the \citet[][Table~1]{cap} ages of 6 galaxy clusters in the redshift range $0.10\le z\le 1.27$. This sample has 
a $1$ Gyr one standard deviation uncertainty
on the age measurements. In all, we have 38 measurements of $t_L^{\rm
obs}(z_i)$ with uncorrelated uncertainties $\sigma_i$. 

For each model and parameter value set ($p$) we compute the $\chi^2$ function
\begin{equation}
\chi^2(p,H_0,t_{\rm inc},t_0^{\rm obs}) = \sum_{i = 1}^{38}\frac{ (t_{\mathrm{L}}(z_i, p,H_0) -
 t_{\mathrm{L}}^{\rm obs}(z_i,t_{\rm inc},t_0^{\rm obs}))^2}{ \sigma_{i}^2 +
\sigma_{t_0^{\rm obs}}^2} + \frac{(t_0(p,H_0)-t_0^{\rm obs})^2}{\sigma_{t^{\rm obs}_0}^2}, 
\label{l}
\end{equation}
\noindent where $\sigma_{t_0}^{\rm obs}$ is the uncertainty in the estimate
of $t_0$ and $t_L(z_i,p)$ and $t_0(p)$ are the predicted values in the model under consideration. From $\chi^2$ we construct a likelihood function $\mathcal{L}'(p,H_0,t_{\rm inc})\propto \exp\left(-\chi^2/2\right)$.

The likelihood function $\mathcal{L}'(p,H_0,t_{\rm inc},t_0^{\rm obs})$ depends on the
total age of the Universe $t_0^{\rm obs}$, incubation time $t_{\rm inc}$ and the Hubble parameter $H_0$. We do not know
$t_{\rm inc}$ and so treat it as a nuisance parameter and analytically
marginalize $\mathcal{L}'$ over it as in \citet{cap,dj}. 
We treat $H_0$ as a nuisance parameter and marginalize over it
with a Gaussian prior with $h=0.742 \pm 0.036$ \citep{riess09}, a little higher
than, but still consistent with, the earlier summary value of 
$h=0.68 \pm 0.04$ \citep{chen03a}. 
We also also apply a bayesian prior as a Gaussian function with central values and variances based on the
WMAP estimate of the total age of the Universe, which is $t_0^{\rm obs}=(13.75\pm
0.13)\ \rm Gyr$ for the $\Lambda$CDM model and $t_0^{\rm
obs}=(13.75^{+0.29}_{-0.27})\ \rm Gyr$ for the XCDM model \citep{dunkley09}.\footnote{The numbers are taken from 
http://lambda.gsfc.nasa.gov/.} 
For
the $\phi$CDM model we assume the same central value as the other two models and
conservatively inflate the error bar to $t_0^{\rm obs}=(13.75\pm0.5)\ \rm Gyr$.
The resulting lookback time likelihood function depends only on the two cosmological
parameters $p$, $\mathcal{L}_L(p)$. The best fit parameters are the pair $p^*$
that maximize the likelihood function and the 1, 2, and 3$\sigma$ confidence
level contours are defined as the sets of cosmological parameters $p_\sigma$
at which the likelihood $\mathcal{L}(p_\sigma)$ is $\exp(-2.30/2)$, $\exp(-6.18/2)$, and $\exp(-11.83/2)$ times smaller than the maximum likelihood $\mathcal{L}(p^*)$. 

To check our method we used the \citet{cap} galaxy cluster data and the
earlier $t_0^{\rm obs}$ result they used and computed the constraints on the XCDM
parameterization. Our contours are consistent with those shown in Fig.~2 of
\citet{cap}. We also used the \citet{svj} passively evolving galaxy ages and
the $t_0^{\rm obs}$ value \citet{dj} used to constrain the $\Lambda$CDM
model. We find that if we pick $h=0.72$ we are able to accurately reproduce
the central and right panels of Fig.~2 of \citet{dj}.

The lookback time versus redshift data constraints on $\Lambda$CDM, XCDM, and
$\phi$CDM are shown in Figs.~1--3.

\section{Results and discussions}

Figure~1 shows the constraints on the $\Lambda$CDM model from the lookback
time and age of the Universe measurements.
The data favor low vales of both $\Omega_{\rm m}$ and $\Omega_\Lambda$ with the best-fit values being
$\Omega_{\rm m}=0.01$ and $\Omega_\Lambda=0.19$. These data prefer spatially-open models, however
a spatially-flat $\Lambda$CDM model with $\Omega_{\rm m}= 0.3$ is less
than 3$\sigma$ from the best fit model. The data constrains $\Omega_{\rm m}$ to be less than 0.45 on 3$\sigma$ confidence level.

Figure~2 presents the constraints on the XCDM parametrization of the equation
of state. The nonrelativistic matter density parameter
is constrained to be less than 0.5 at 3$\sigma$ confidence. Low values of
$\Omega_{\rm m}$ are favored 
with the best-fit values being $\Omega_{\rm m}=0.03$ and $\omega_{\rm x}=-0.41$ 
and a spatially-flat model with $\Omega_{\rm m}= 0.3$ is about
2$\sigma$ from the best fit model.

Figure~3 shows the constraints on the $\phi$CDM model of dark energy. In this
model the nonrelativistic matter density parameter is less than
0.5 at 3$\sigma$ confidence. The $\alpha$ parameter on the other hand is not well constrained. The best-fit parameter value is
$\alpha=10$, but the likelihood is very flat in the direction of $\alpha$ and the difference between the best-fit value and $\alpha=0$ (which is the spatially-flat $\Lambda$CDM case)
is slightly less than 2$\sigma$.

Current lookback time data by themselves are unable to tightly constrain cosmological
parameters. Constraints from galaxy cluster gas mass fraction versus redshift
data \citep[e.g.,][]{chen04}, SNeIa apparent magnitude versus redshift
measurements \citep[e.g.,][]{wilson06}, and baryon acoustic peak data
\citep[e.g.,][]{samushia09b} are more restrictive than the lookback time
constraints. However, the constraints from lookback time data are somewhat tighter than the constraints from strong gravitational
lensing data \citep[e.g.,][]{chae04}, measurements of the Hubble parameter as a function
of redshift \citep[e.g.,][]{samushia07}, radio galaxy angular size versus redshift data \citep[e.g.,][]{daly09},
and gamma-ray burst luminosity distance versus redshift data \citep[e.g.,][]{samushia09c}.

To get tighter constraints on cosmological parameters we combine the lookback time data and the measurement of the age of the Universe
with baryon acoustic peak data \citep{percivaletal07} and SNeIa
``Union'' apparent magnitude versus redshift measurements \citep{kowalski09}. Since these data sets are independent we
compute a joint likelihood function that is a product of individual
likelihood functions
\begin{equation}
\mathcal{L}_{\rm joint}=\mathcal{L}_{L}\mathcal{L}_{\rm BAO}\mathcal{L}_{\rm
SNe},
\end{equation}
\noindent and define the best-fit parameters and confidence level contours as
discussed above.

The constraints on the three dark energy models from a joint analysis of
these data are
shown in Figs.~4--6. Currently available lookback time data do not significantly change the results derived using BAO peak measurements and
SNeIa apparent magnitude data. In all three dark energy models when lookback
time data are added to the mix the confidence level regions favor slightly
smaller 
values of nonrelativistic matter density parameter $\Omega_{\rm m}$.

Overall, current data is a good fit to all three dark energy models. For
$\phi$CDM and XCDM they slightly favor time-dependent
dark energy, but the time-independent cosmological constant is also a good
fit.

We anticipate that a new, improved data set of lookback times will soon be
available (R.~Jimenez, private communication, 2009.) With more and better data 
we expect significantly tighter constraints on dark energy parameters. The
lookback time versus redshift test, either by itself or at least in
combination with other cosmological probes,
could prove very useful in detecting or constraining dark energy time
evolution.

\section{Acknowledgements}
Abha Dev and Deepak Jain thank A.~Mukherjee and S. Mahajan for providing
facilities to carry out this research. Deepak Jain acknowledges the hospitality
provided by the IUCAA, Pune where the part of the work is done. DJ and AD
acknowledge the financial support provided by Department of Science and
Technology, Govt. of India under project No. SR/S2/HEP-002/2008. We acknowledge
support from DOE grant DE-FG03-99EP41093, Georgian National Science foundation
grant ST08/4-442, and Scientic Co-operation Programme between Eastern Europe and
Switzerland (SCOPES) grant.

\begin{figure}
\includegraphics[width=150mm, height=150mm]{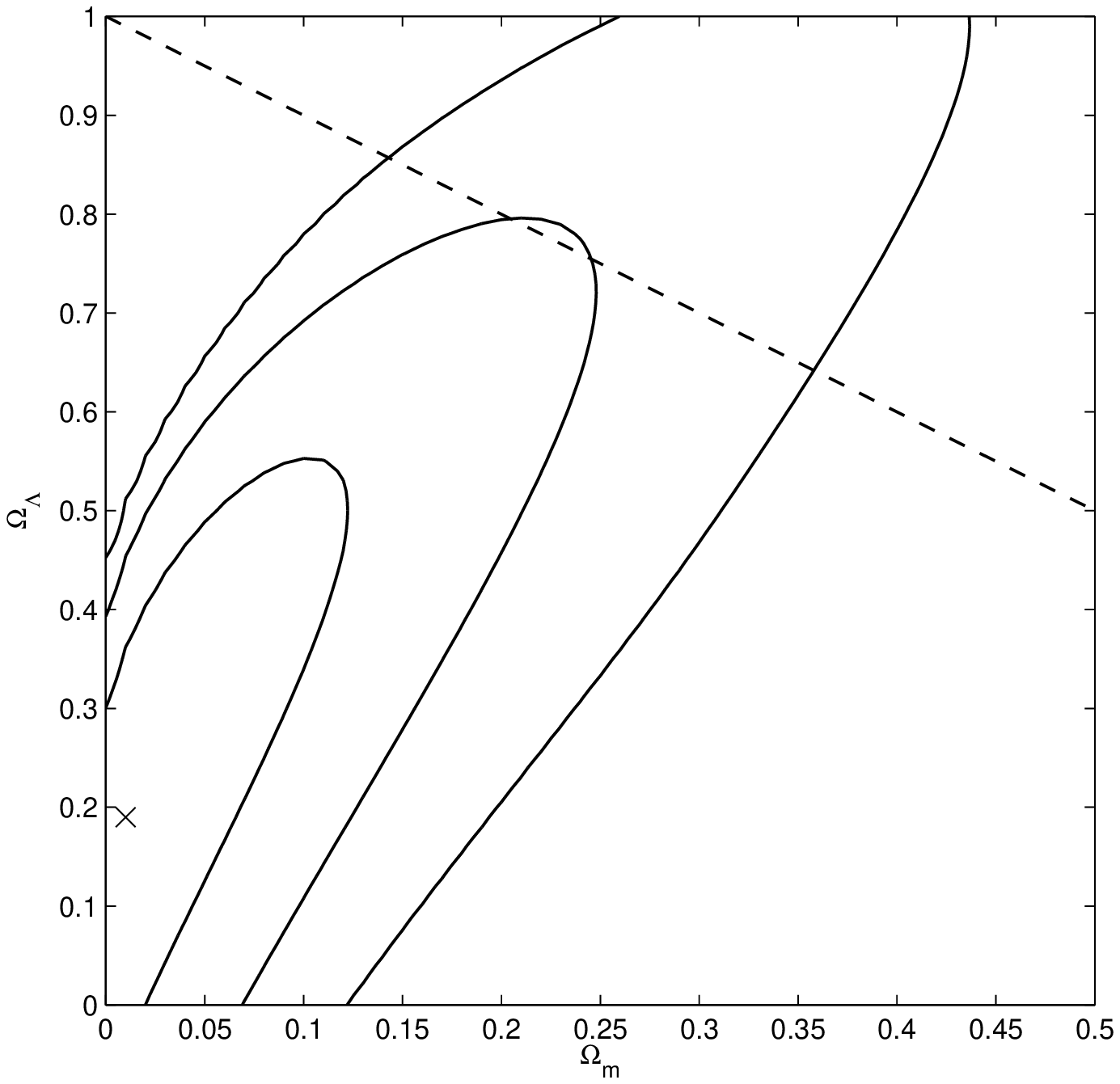}
\caption{1, 2, and 3$\sigma$ confidence level contours for the $\Lambda$CDM
model from the lookback time data and measurement of the age of the Universe.
The dashed line corresponds to spatially-flat models. The cross indicates the
best-fit parameters $\Omega_{\rm m}=0.01$ and $\Omega_\Lambda=0.19$ with
$\chi^2=33$ for 37 degrees of freedom.}
\end{figure}

\begin{figure}
\includegraphics[width=150mm, height=150mm]{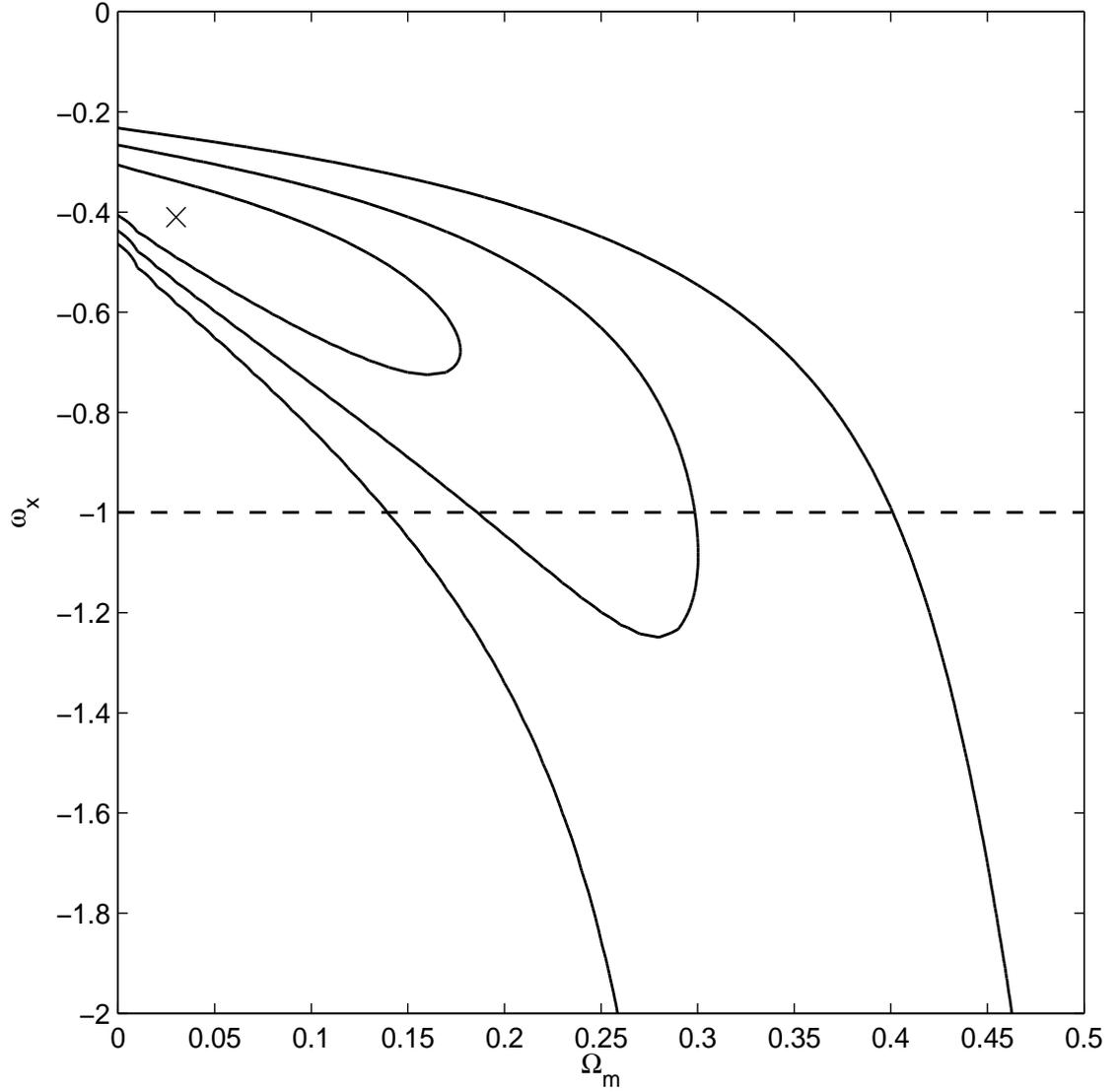}
\caption{1, 2, and 3$\sigma$ confidence level contours for the XCDM
parameterization of dark energy in a spatially-flat cosmological model, from the lookback time data and measurement of the age
of the Universe. The dashed $\omega_{\rm x}=-1$ line corresponds to 
spatially-flat $\Lambda$CDM models. The cross indicates the best-fit parameters $\Omega_{\rm
m}=0.03$ and $\omega_{\rm x}=-0.41$ with $\chi^2=28$ for 37 degrees of freedom.}
\end{figure}

\begin{figure}
\includegraphics[width=150mm, height=150mm]{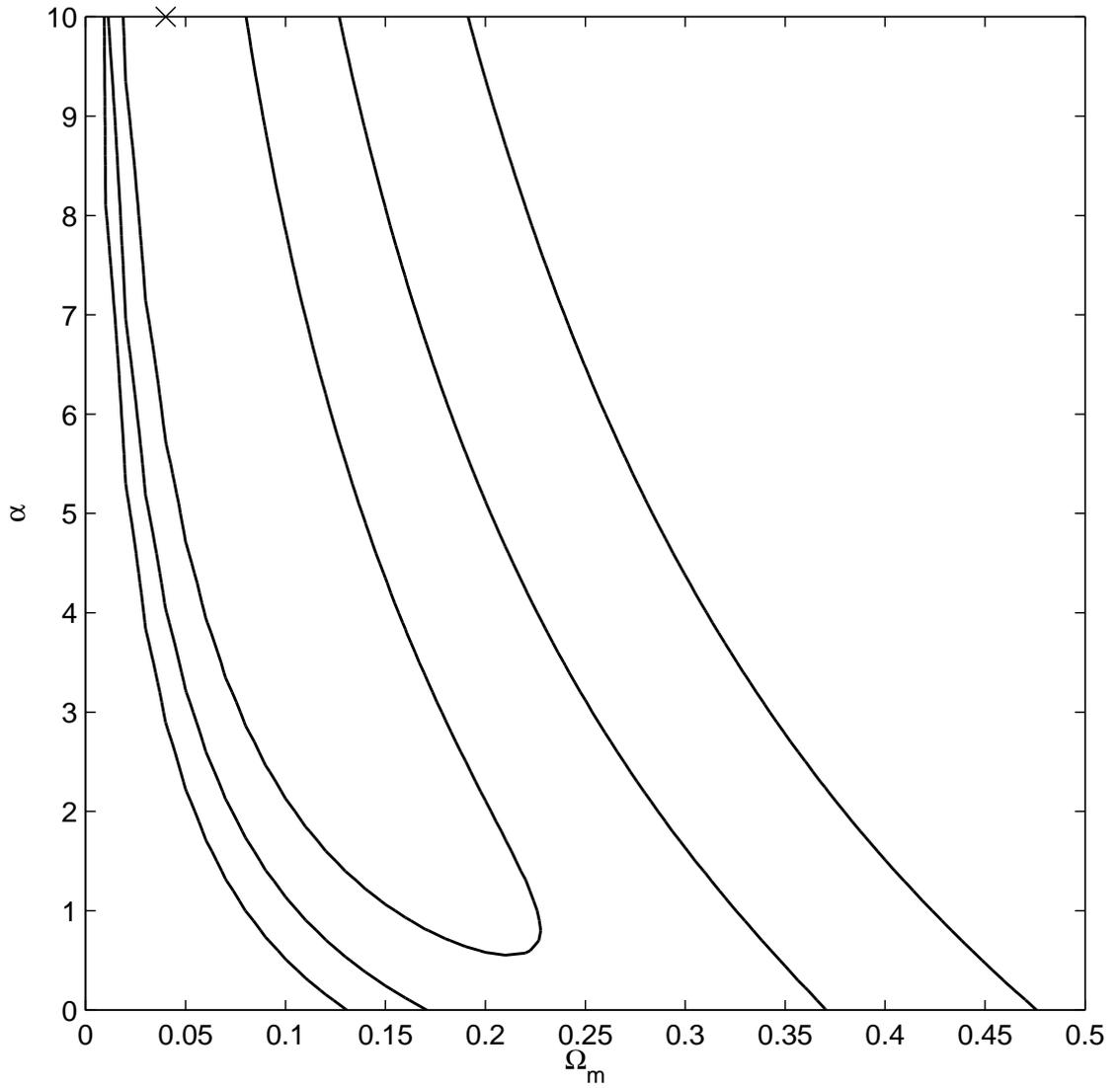}
\caption{1, 2, and 3$\sigma$ confidence level contours for the spatially-flat $\phi$CDM
model from the lookback time data and measurement of the age of the Universe.
The $\alpha=0$ horizontal axis corresponds to spatially-flat $\Lambda$CDM
models. The
cross indicates the best-fit parameters $\Omega_{\rm m}=0.04$ and
$\alpha=10$ with $\chi^2=22$ for 37 degrees of freedom.}
\end{figure}

\begin{figure}
\includegraphics[width=150mm, height=150mm]{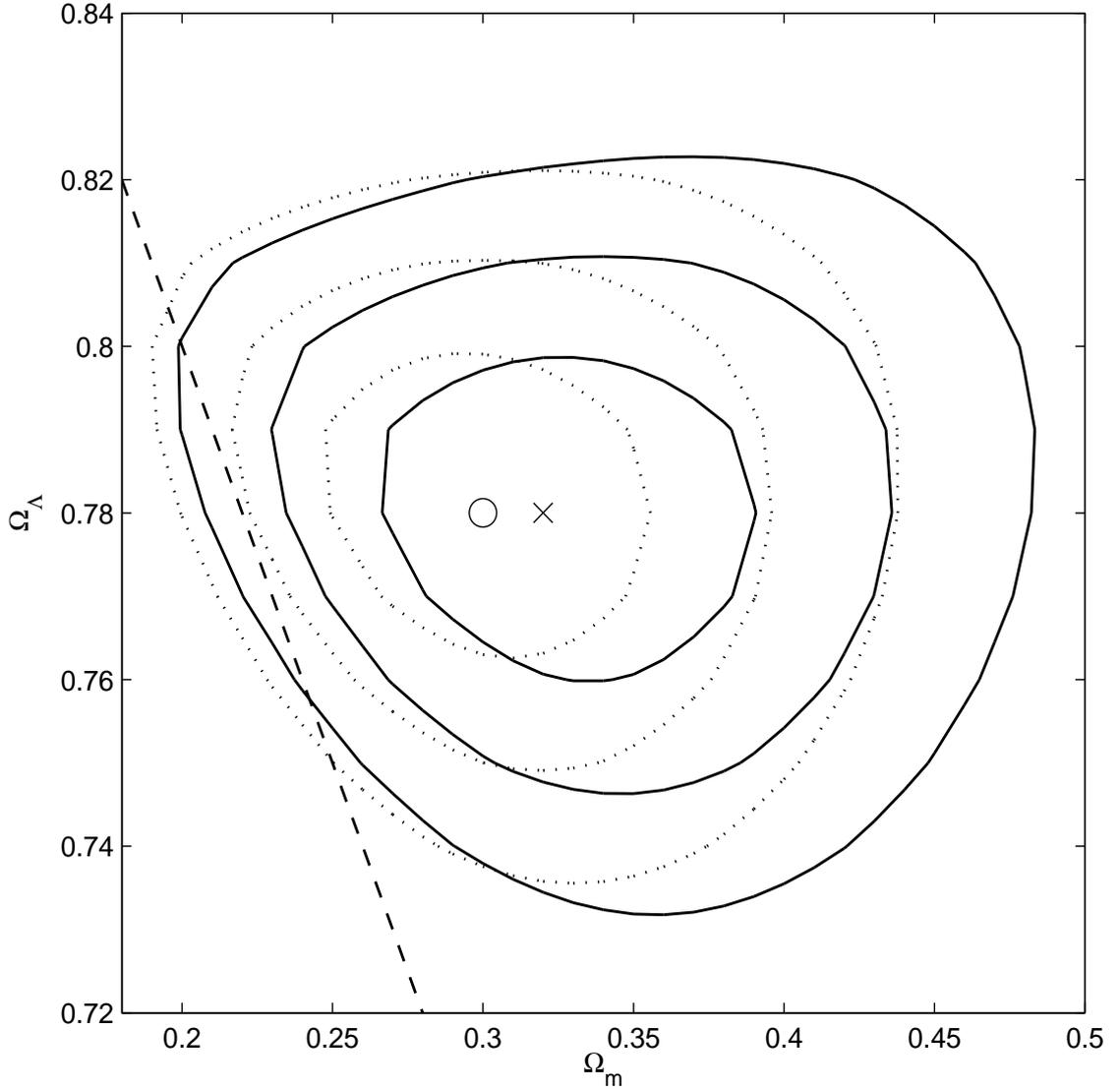}
\caption{1, 2, and 3$\sigma$ confidence level contours for the $\Lambda$CDM
model. Numerical noise is responsible for the jaggedness of parts of the
contours. The dashed line demarcates spatially-flat models. Dotted lines
(circle denotes the best-fit point at $\Omega_{\rm m}=0.30$ and
$\Omega_\Lambda=0.78$ with $\chi^2=359$ for 346 degrees of freedom) are
derived using the lookback time data, measurement of the age of the Universe,
SNeIa Union data, and BAO peak measurements, while solid lines (cross
denotes the best-fit point at $\Omega_{\rm m}=0.32$ and
$\Omega_\Lambda=0.78$ with $\chi^2=318$ for 307 degrees of freedom) are derived
using SNeIa and BAO data. The dashed line corresponds to spatially-flat
 models.}
\end{figure}

\begin{figure}
\includegraphics[width=150mm, height=150mm]{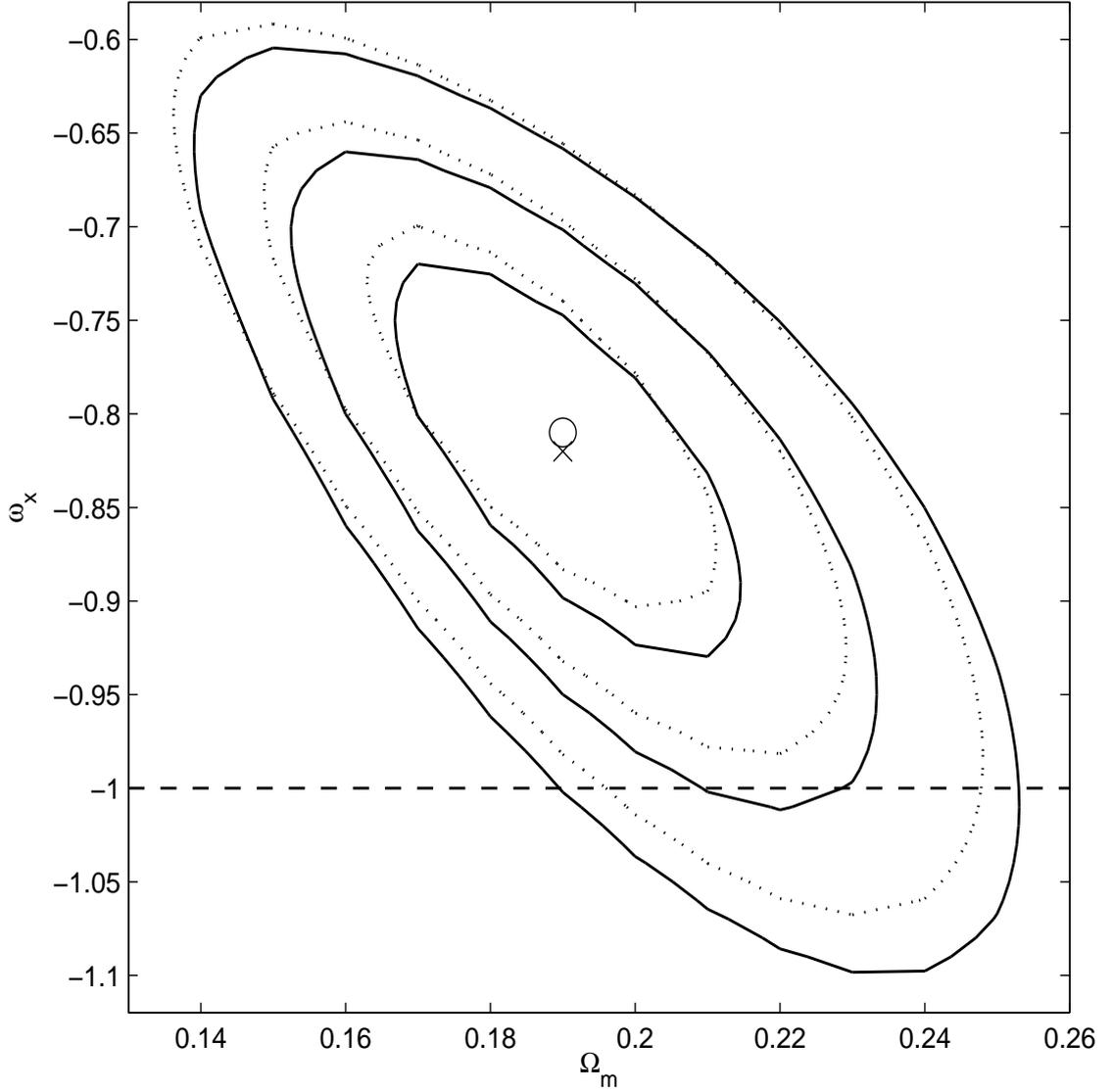}
\caption{1, 2, and 3$\sigma$ confidence level contours for the XCDM
parameterization of dark energy in a spatially-flat cosmological model. The dashed line demarcates spatially-flat $\Lambda$CDM
models. Dotted lines (circle denotes the best-fit point at $\Omega_{\rm
m}=0.19$ and $\omega_{\rm x}=-0.80$ with $\chi^2=352$ for 346 degrees of
freedom) are derived using the lookback time data, measurement of the age of
the Universe, SNeIa Union data, and BAO peak measurements, while solid lines
(cross denotes the best-fit point at $\Omega_{\rm m}=0.19$ and $\omega_{\rm
x}=-0.81$ with $\chi^2=321$ for 307 degrees of freedom) are derived using
only SNeIa and BAO data. The dashed $\omega_{\rm x}$ line corresponds to
spatially-flat $\Lambda$CDM models.}
\end{figure}

\begin{figure}
\includegraphics[width=150mm, height=150mm]{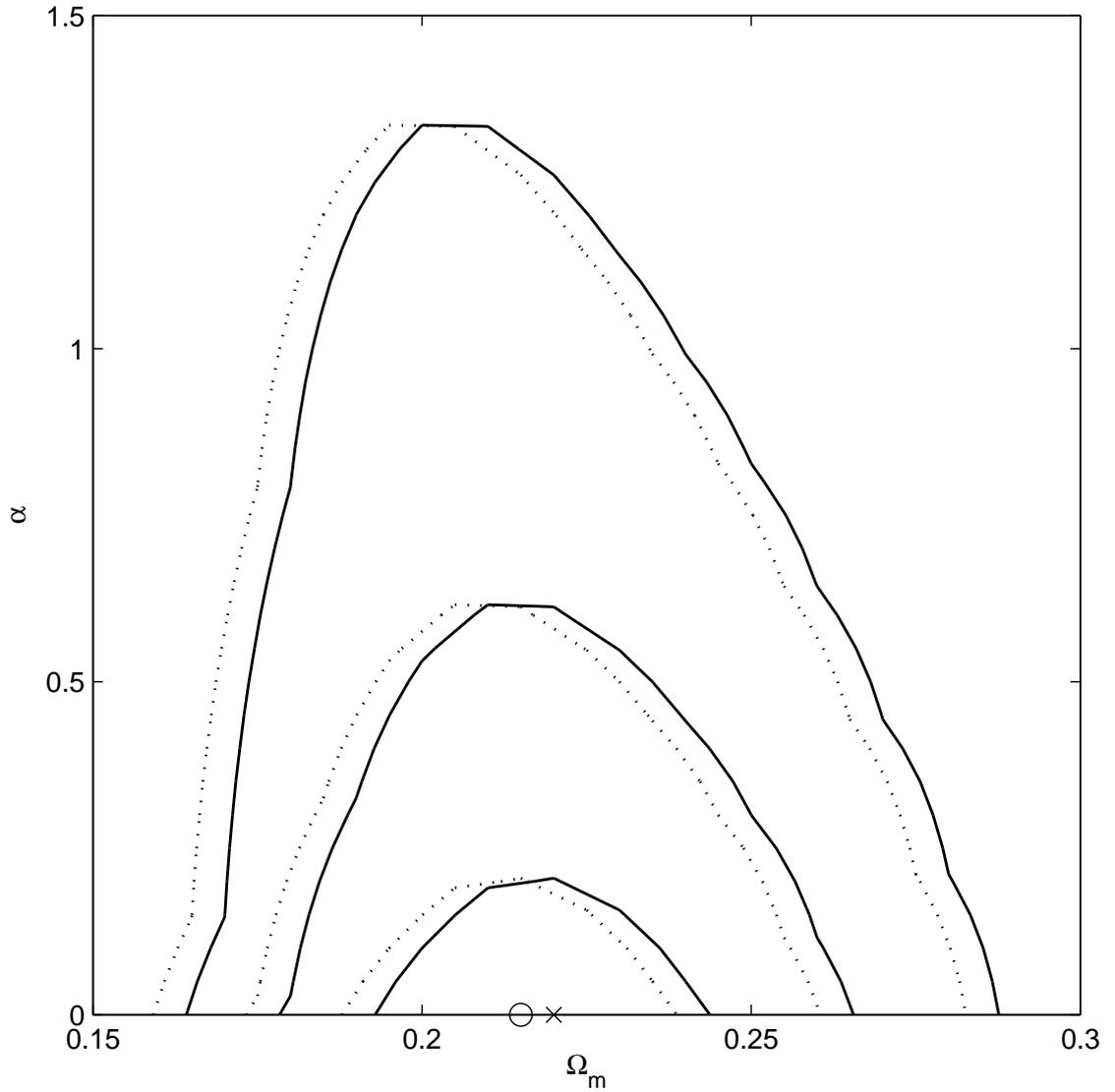}
\caption{1, 2, and 3$\sigma$ confidence level contours for the spatially-flat $\phi$CDM
model. The $\alpha=0$ horizontal axis corresponds to spatially-flat
$\Lambda$CDM models. Dotted lines (circle denotes the best-fit point at
$\Omega_{\rm m}=0.215$ and $\alpha=0.0$ with $\chi^2=359$ for 346 degrees of
freedom) are derived using the lookback time data, measurement of the age of
the Universe, SNeIa Union data, and BAO peak measurements, while solid lines
(cross denotes the best-fit point at $\Omega_{\rm m}=0.22$ and $\alpha=0.0$
with $\chi^2=329$ for 307 degrees of freedom) are derived using only SNeIa and BAO data.}
\end{figure}

\clearpage

\begin{tabular}{|c|c|}
\hline
\multicolumn{2}{|c|}{\citet{svj} galaxy ages}\\
\hline
$z_i$ & $t_i(z_i)$ (Gyr)\\
\hline
0.1171 & 10.2  \\  
0.1174 &  10.0 \\ 
0.2220 &   9.0 \\ 
0.2311 &   9.0 \\
0.3559 &   7.6 \\
0.4520 &   6.8 \\  
  0.5750 &  7.0\\
0.6440 &  6.0\\
0.6760 &  6.0\\
0.8330 &  6.0\\
0.8360 &  5.8\\
0.9220 &  5.5\\
1.179 &  4.6\\
  1.222 & 3.5 \\
1.224 &   4.3 \\ 
1.225 &  3.5 \\
1.226 &  3.5 \\ 
1.340 &  3.4 \\ 
1.380 &  3.5 \\  
1.383 &  3.5 \\  
1.396 &  3.6 \\  
  1.430 &  3.2 \\ 
1.450 &   3.2 \\  
1.488 &  3.0 \\  
1.490 &  3.6 \\  
1.493 &  3.2 \\  
1.510 &  2.8 \\  
1.550 &  3.0 \\  
1.576 &  2.5 \\ 
  1.642 &  3.0 \\  
1.725 &    2.6 \\  
1.845 & 2.5 \\  
\hline
\end{tabular}

\end{document}